\documentclass[preprint,superscriptaddress,preprintnumbers,showpacs]{elsart}
\usepackage{amssymb}
\usepackage{dcolumn}
\usepackage{bm}
\usepackage{graphicx}
\usepackage{booktabs}
\usepackage{multirow}

\journal{Physics Letter B. Accepted for publication.}

\begin{document}

\begin{frontmatter}

\title{A prediction of neutrino mixing matrix with CP violating phase}

\author[1]{Xinyi Zhang},
\author[1,2]{Bo-Qiang Ma\corauthref{*}} \corauth[*]{Corresponding author.}\ead{mabq@pku.edu.cn}
\address[1]{School of Physics and State Key Laboratory of Nuclear Physics and
Technology, Peking University, Beijing 100871, China}
\address[2]{Center for High Energy
Physics, Peking University, Beijing 100871, China}

\begin{abstract}
The latest experimental progress have established three kinds of
neutrino oscillations with three mixing angles measured to rather
high precision. There is still one parameter, i.e., the CP violating
phase, missing in the neutrino mixing matrix. It is shown that a
replay between different parametrizations of the mixing matrix can
determine the full neutrino mixing matrix together with the CP
violating phase. From the maximal CP violation observed in the
original Kobayashi-Maskawa (KM) scheme of quark mixing matrix, we
make an Ansatz of maximal CP violation in the neutrino mixing
matrix. This leads to the prediction of all nine elements of the
neutrino mixing matrix and also a remarkable prediction of the CP
violating phase $\delta_{\rm CK}=(85.48^{+4.67(+12.87)}_{-1.80(-4.90)})^\circ$ within
$1\sigma~(3\sigma)$ range from available experimental information. We also
predict the three angles of the unitarity triangle corresponding to the quark sector for confronting with
the CP-violation related measurements.
\end{abstract}


\end{frontmatter}
\newpage

The recent measurements of the neutrino mixing angle $\theta_{13}$
by the T2K, MINOS and Double Chooz collaborations~\cite{newexpt},
especially the latest ones by the Daya Bay
Collaboration~\cite{Daya-Bay} and the RENO
Collaboration~\cite{Reno}, have led to the establishment of three
kinds of neutrino oscillations. The three mixing angles, i.e.,
$\theta_{12}$, $\theta_{23}$, and $\theta_{13}$, have been measured
to rather high precision, and there have been some
perspectives~\cite{Zhang:2012xu,Li:2011ag,Zhang:2012zh,Xing:2012ej,Wu:2012ri,Meloni:2012sy}
by these novel experimental progress.
As the three mixing angles are
sizable, the neutrino physics has entered an era of precise
measurement. A promising chance is viable for the measurement of the
CP violating phase $\delta$ in future experiments. It is thus timely
to look at the CP violating phase from theoretical
aspects~\cite{Xing:2012ej,Wu:2012ri,Meloni:2012sy}.

The mixing of fermions is a significant feature of fundamental
particles, i.e., of quarks and leptons, and the mixing is well
described by fermion mixing matrices~\cite{PMNS,Cabibbo,KM}. The
mixing of quarks is described by the Cabibbo-Kobayashi-Maskawa (CKM)
matrix~\cite{Cabibbo,KM}, with all parameters, i.e., three mixing
angles and one CP violating phase, determined to rather high
precision experimentally. The misalignment of the flavor eigenstates
with the mass eigenstates in the lepton sector is also described by
a mixing matrix, namely the Pontecorvo-Maki-Nakagawa-Sakata (PMNS)
matrix~\cite{PMNS}. The PMNS matrix is defined as $U_{\rm
PMNS}=U^{l\dagger}_LU^\nu_L$ and can be expressed generally as
\begin{eqnarray}
 U_{\rm PMNS}=\left(
  \begin{array}{ccc}
    U_{e1}    & U_{e2}    & U_{e3}   \\
    U_{\mu1}  & U_{\mu2}  & U_{\mu3} \\
    U_{\tau1} & U_{\tau2} & U_{\tau3}\\
  \end{array}\right).
\end{eqnarray}
In the representation that the mass matrix of charged leptons is
diagonal, the PMNS matrix represents the neutrino mixing, therefore
we can also call it the neutrino mixing matrix. In case the
neutrinos are of Dirac type, the neutrino mixing matrix can be
parameterized by three rotation angles and a CP violating phase. Two
additional phase angles are needed for the PMNS matrix if the
neutrinos are of Majorana type. For the neutrino mixing, the
Majorana phase angles do not affect the absolute values of the
elements of mixing matrix and are omitted in the following
discussion.

There are many possible ways to parameterize the mixing matrix in
terms of four independent parameters. One of such parametrizations
is the Chau-Keung (CK) scheme~\cite{CK} adopted by Particle Data
Group~\cite{PDG1996,PDG2008,pdg2010} as the standard one, which is
\begin{eqnarray}
U_{\rm CK}&=&\left(
  \begin{array}{ccc}
    1  & 0     & 0         \\
    0  & c_{23}  & s_{23} \\
    0  & -s_{23} & c_{23} \\
  \end{array}
\right)\left(
  \begin{array}{ccc}
    c_{13}                & 0 & s_{13}e^{-i\delta_{\rm CK}} \\
    0                     & 1 & 0          \\
    -s_{13}e^{i\delta_{\rm CK}} & 0 & c_{13} \\
  \end{array}
\right)\left(
  \begin{array}{ccc}
    c_{12}  & s_{12} & 0 \\
    -s_{12} & c_{12} & 0 \\
    0       & 0      & 1 \\
  \end{array}
\right)\nonumber\\
&=&\left(
\begin{array}{ccc}
c_{12}c_{13} & s_{12}c_{13} & s_{13}e^{-i\delta_{\rm CK}}         \\
-s_{12}c_{23}-c_{12}s_{23}s_{13}e^{i\delta_{\rm CK}} & c_{12}c_{23}-s_{12}s_{23}s_{13}e^{i\delta_{\rm CK}} & s_{23}c_{13} \\
s_{12}s_{23}-c_{12}c_{23}s_{13}e^{i\delta_{\rm CK}} & -c_{12}s_{23}-s_{12}c_{23}s_{13}e^{i\delta_{\rm CK}} & c_{23}c_{13}\\
\end{array}
\right).
\end{eqnarray}
Another well-discussed parametrization is the original
Kobayashi-Maskawa (KM) scheme~\cite{KM}, which is,
\begin{eqnarray}
U_{\rm KM}&=&\left(
  \begin{array}{ccc}
    1 & 0   & 0    \\
    0 & c_2 & -s_2 \\
    0 & s_2 & c_2  \\
  \end{array}
\right)\left(
  \begin{array}{ccc}
    c_1 & -s_1 & 0 \\
    s_1 & c_1  & 0 \\
    0   & 0    & e^{i\delta_{\rm KM}} \\
  \end{array}
\right)\left(
  \begin{array}{ccc}
    1 & 0   & 0    \\
    0 & c_3 & s_3  \\
    0 & s_3 & -c_3 \\
  \end{array}
\right)\nonumber\\
&=&\left(
\begin{array}{ccc}
c_1     & -s_1c_3                     & -s_1s_3           \\
s_1c_2  & c_1c_2c_3-s_2s_3e^{i\delta_{\rm KM}} & c_1c_2s_3+s_2c_3e^{i\delta_{\rm KM}} \\
s_1s_2  & c_1s_2c_3+c_2s_3e^{i\delta_{\rm KM}} &
c_1s_2s_3-c_2c_3e^{i\delta_{\rm KM}}
\end{array}\label{KM}
\right).
\end{eqnarray}
It is rather interesting that this scheme allows for almost perfect
maximal CP violation, i.e., the CP violating phase
$\delta^{\mathrm{quark}}_{\mathrm{KM}}=90^\circ$, for
quarks~\cite{koide,boomerang,Li:2010ae,qinnan,Ahn:2011it}, whereas
in the standard parametrziation
$\delta^{\mathrm{quark}}_{\mathrm{CK}}=68.9^\circ$\cite{pdg2010},
which deviates from the maximal CP violation. This inspires us to
make a prediction of the neutrino mixing matrix with all nine
elements determined based on experimental information of three
mixing angles together with an Ansatz of maximal CP violation for
the KM-scheme of mixing matrix.

We should notice that the absolute values of the corresponding
elements of the mixing matrix should be the same for different
parametrizations, but the phase of each element may differ
significantly. Also the degree of CP violation, such as whether it
is maximal or minimal, is parametrization dependent. Most previous
information of neutrino mixing matrix are expressed by parameters in
the standard parametrization, and we still cannot combine the three
measured mixing angles with the maximal CP phase in the KM-scheme in
a direct way to predict the nine elements of the mixing matrix. It
is thus necessary to make a replay between different schemes for a
full prediction of the nine elements of the neutrino mixing matrix
together with the CP violating phase $\delta_{\mathrm{CK}}$ in the
standard parametrization.

The observables of the neutrino oscillation experiments are related
to the mixing angles of the standard parametrization. A global
fitting of neutrino mixing angles based on previous experimental
data and T2K and MINOS
experiments~($1\sigma~(3\sigma)$)~\cite{global} gives,\\
\begin{eqnarray}
&\sin^2\theta_{12}=0.312,\quad0.296-0.329(1\sigma),\quad0.265-0.364(3\sigma);\label{input1}\\
&\sin^2\theta_{23}=0.42,\quad0.39-0.50(1\sigma),\quad0.34-0.64(3\sigma).
\end{eqnarray}
Combined with the latest result
\begin{eqnarray}
\sin^2\theta_{13}=0.024,\quad0.020-0.028(1\sigma),\quad0.010-0.038(3\sigma)\label{input2}
\end{eqnarray}
from the Daya Bay Collaboration~\cite{Daya-Bay}, we can get five
moduli of the PMNS matrix elements from the standard parametrization
without knowledge of the CP violating phase. Notice that the error
range for $\theta_{13}$ of the Daya Bay result is calculated by an
assumption of Gaussian distribution, and the $1\sigma$ deviation is
estimated by $\sigma^2=\sigma_{\mathrm{stat}}^2 +
\sigma_{\mathrm{syst}}^2$. The five matrix elements are,
\begin{eqnarray}
&|U_{e1}|=c_{12}c_{13}=\sqrt{(1-s^2_{12})(1-s^2_{13})}=0.8195^{+0.010(+0.032)}_{-0.010(-0.029)};\\
&|U_{e2}|=s_{12}c_{13}=\sqrt{s^2_{12}(1-s^2_{13})}=0.5518^{+0.015(+0.046)}_{-0.014(-0.042)};\\
&|U_{e3}|=|s_{13}|=\sqrt{s^2_{13}}=0.1549\pm0.013(\pm0.045);\\
&|U_{\mu3}|=s_{23}c_{13}=\sqrt{s^2_{23}(1-s^2_{13})}=0.6403^{+0.061(+0.168)}_{-0.023(-0.061)};\\
&|U_{\tau3}|=c_{23}c_{13}=\sqrt{(1-s^2_{23})(1-s^2_{13})}=0.7524^{+0.052(+0.143)}_{-0.020(-0.052)}.
\end{eqnarray}
With these five moduli, together with an Ansatz of maximal CP
violation $\delta_{\mathrm{KM}}=90^\circ$, we can get the mixing
angles in the KM parametrization, which are,
\begin{eqnarray}
\theta_1=(34.97^{+1.00(+3.20)}_{-1.00(-2.90)})^\circ,\quad\theta_2=(39.87^{+5.18(+14.21)}_{-1.97(-5.18)})^\circ,\quad\theta_3=(15.68^{+1.34(+4.63)}_{-1.33(-4.60)})^\circ.
~~~~~~
\end{eqnarray}
The corresponding trigonometric functions are,
\begin{eqnarray}
&\sin\theta_1=0.5732^{+0.014(+0.046)}_{-0.014(-0.042)},\quad\cos\theta_1=0.8194^{+0.010(+0.032)}_{-0.010(-0.029)};\\~~~~~~~~~~
&\sin\theta_2=0.6411^{+0.069(+0.190)}_{-0.026(-0.069)},\quad\cos\theta_2=0.7674^{+0.058(+0.159)}_{-0.022(-0.058)};\\~~~~~~~~~~
&\sin\theta_3=0.2703\pm0.021(\pm0.078),\quad\cos\theta_3=0.9628\pm0.006(\pm0.022).\label{KMpara}~~~~~~~~~~
\end{eqnarray}
Combined with the maximal CP violating phase
$\delta_{\mathrm{KM}}=90^\circ$ and the trigonometric functions in
Eq.(\ref{KM}), we can get all the moduli of the PMNS matrix, which
is,
\begin{eqnarray}
|U_{\rm PMNS}|= \left(
  \begin{array}{ccc}
   0.8195^{+0.010(+0.032)}_{-0.010(-0.029)}& 0.5518^{+0.015(+0.046)}_{-0.014(-0.042)}  &
   0.1549\pm0.013(\pm0.045)   \\
   0.4399^{+0.045(+0.129)}_{-0.024(-0.068)}& 0.6297^{+0.045(+0.123)}_{-0.018(-0.048)}  &  0.6403^{+0.061(+0.168)}_{-0.023(-0.061)}  \\
   0.3675^{+0.049(+0.140)}_{-0.024(-0.068)}& 0.5467^{+0.051(+0.143)}_{-0.021(-0.059)}  &  0.7524^{+0.052(+0.143)}_{-0.020(-0.052)}
  \end{array} \right).~~~~~~\label{PMNSprediction}
\end{eqnarray}
Then we can work out the CP violating phase in the standard
parametrization, using the following expressions,
\begin{eqnarray}
&|U_{\mu1}|=|-s_{12}c_{23}-c_{12}s_{23}s_{13}e^{i\delta_{\rm CK}}|;\\
&|U_{\mu2}|=|c_{12}c_{23}-s_{12}s_{23}s_{13}e^{i\delta_{\rm CK}}|;\\
&|U_{\tau1}|=|s_{12}s_{23}-c_{12}c_{23}s_{13}e^{i\delta_{\rm CK}}|;\\
&|U_{\tau2}|=|-c_{12}s_{23}-s_{12}c_{23}s_{13}e^{i\delta_{\rm CK}}|.
\end{eqnarray}
We can calculate $\delta_{\rm CK}$ from one of the above four
equations. Using the original input from Eq.~(\ref{input1}) to
Eq.~(\ref{input2}) in the process of calculating, we get the same
central values of $\delta_{\rm CK}$ and the same error bars from the
four equations given the effective digits we keep in the result. The
resulting $\delta_{\rm CK}$ from the above expressions is
\begin{equation}
\delta_{\rm CK}=(85.48^{+4.67(+12.87)}_{-1.80(-4.90)})^\circ
\label{CPprediction}
\end{equation}
within $1\sigma~(3\sigma)$ range. The four elements of $|U_{\mu1}|$, $|U_{\mu2}|$,
$|U_{\tau1}|$ and $|U_{\tau2}|$ at the left-lower corner of
Eq.~(\ref{PMNSprediction}) are predictions of our analysis. The full
neutrino mixing matrix predicted in Eq.~(\ref{PMNSprediction}) can
be used to construct also the phase factors for all nine elements
once a specific scheme of parametrization is
chosen~\cite{Zhang:2012xu,Zhang:2012zh}. Our predictions of the full
neutrino mixing matrix Eq.~(\ref{PMNSprediction}) together with the
CP violating phase Eq.~(\ref{CPprediction}) in the standard
parametrization can be conveniently applied for phenomenological
analysis.

By the way, it is helpful to work out the Jarlskog
invariant~\cite{jarlskog} in the two schemes of parametrizations
above,
\begin{eqnarray}
&\mathcal{J}_{\rm CK}=\frac{1}{8}\cos\theta_{13}\sin2\theta_{12}\sin2\theta_{23}\sin2\theta_{13}\sin\delta_{\rm CK}=0.0345^{+0.0029(+0.0100)}_{-0.0028(-0.0097)};~~~~~~~~~\\
&\mathcal{J}_{\rm
KM}=\frac{1}{8}\sin\theta_1\sin2\theta_1\sin2\theta_2\sin2\theta_3\sin\delta_{\rm
KM}=0.0345^{+0.0030(+0.0101)}_{-0.0028(-0.0097)},~~~~~~~~~
\end{eqnarray}
which are consistent with each other. The value of this parameter is
sizable than previous expectation and it is thus meaningful to
design experiments for the measurement of the CP violating phase
through neutrino oscillation processes. We notice that
$\mathcal{J}_{\rm KM}$ possesses a maximal CP violation as implied
from our Ansatz $\delta_{\rm KM}=90^\circ$, whereas
$\mathcal{J}_{\rm CK}$ is close to a maximal CP violation as a
phenomenological consequence from our analysis.

The unitarity triangles constructed from the unitarity conditions
$\Sigma_i U_{ij}U_{ik}^*=\delta_{jk}(j\neq k)$ and $\Sigma_j
U_{ij}U_{kj}^*=\delta_{ik}(i\neq k)$ carry information on the CP
violation~\cite{Koide:2006fr,Li:2012zx}. Actually, in the quark
sector, the CP violating information can be obtained from the
observables $\alpha$, $\beta$, $\gamma$, which are the inner angles
of the $db$ unitarity triangle. As is pointed out in
Ref.~\cite{Farzan:2002ct}, the possibility of reconstructing the
unitarity triangle can be viewed as an alternative way in search for
the CP violation in both the oscillation and nonoscillation
experiments. It is worth mention that the unitarity triangles carry
information of CP violation in a convention-independent way, and
this makes them a better candidate in comparison with the CP
violating phase $\delta$ as in any angle-phase parametrizations. As
a result, it is worthwhile to calculate the inner angles of the
$\nu_2 \nu_3$ unitarity triangle,
\begin{eqnarray}
U_{e2}U_{e3}^*+U_{\mu2}U_{\mu3}^*+U_{\tau2}U_{\tau3}^*=0,
\end{eqnarray}
which is the correspondent of the $db$ unitarity triangle in the lepton sector~\cite{Bjorken:2005rm}. The result is,
\begin{eqnarray}
\alpha=\varphi_2=\arg(-\frac{U_{\tau2}U_{\tau3}^*}{U_{e2}U_{e3}^*})=(78.58^{+4.15(+11.63)}_{-1.79(-5.28)})^\circ;\\
\beta=\varphi_1=\arg(-\frac{U_{\mu2}U_{\mu3}^*}{U_{\tau2}U_{\tau3}^*})=(12.00^{+1.84(+5.65)}_{-1.22(-4.10)})^\circ;\\
\gamma=\varphi_3=\arg(-\frac{U_{e2}U_{e3}^*}{U_{\mu2}U_{\mu3}^*})=(89.42^{+3.94(+10.85)}_{-1.53(-4.06)})^\circ.
\end{eqnarray}
As the unitarity triangle is convention-independent,
we adopt the KM scheme parameters in Eq.~[\ref{KMpara}] together
with our Ansatz $\delta_{\rm KM}=90^\circ$ as input for the complex PMNS matrix.
The above result can be tested when the unitarity triangles can be reconstructed from
future experiments related to the CP-violation of neutrinos.

For the $\nu_e$ appearance channel, i.e., one of the golden channels for leptonic CP violation, the oscillation probability is~\cite{freund},
\begin{eqnarray}
&&P(\nu_{\mu} \rightarrow \nu_e) \approx \sin^2 \theta_{23} {\sin^2
2
\theta_{13}\over (\hat{A}-1)^2}\sin^2((\hat{A}-1)\Delta) \nonumber\\
&& +\alpha{\sin\delta_{CP}\cos\theta_{13}\sin 2 \theta_{12} \sin 2
\theta_{13}\sin 2 \theta_{23}\over \hat{A}(1-\hat{A})}
\sin(\Delta)\sin(\hat{A}\Delta)\sin((1-\hat{A})\Delta) \nonumber\\
&& +\alpha{\cos\delta_{CP}\cos\theta_{13}\sin 2 \theta_{12} \sin 2
\theta_{13}\sin 2 \theta_{23}\over \hat{A}(1-\hat{A})}
\cos(\Delta)\sin(\hat{A}\Delta)\sin((1-\hat{A})\Delta) \nonumber\\
&& +\alpha^2 {\cos^2\theta_{23}\sin^2 2 \theta_{12}\over
\hat{A}^2}\sin^2(\hat{A}\Delta),  \label{qe1}
\end{eqnarray}
where $\alpha=\Delta m^2_{21}/\Delta m^2_{31}$, $\Delta = \Delta
m^2_{31} L/4E$, $\hat{A}=2 V E/\Delta m^2_{31}$, and $V=\sqrt{2} G_F
n_e$. $n_e$ is the density of electrons in the Earth and $\hat{A}$
describes the strength of the matter effects. Measurements of this
probability over different beam energies can impose constraints in
the ($\theta_{13}$, $\delta_{\rm CP}$) parameter space. It is
pointed in Ref.~\cite{Bishai:2012ss} that combining 8~GeV and 60~GeV
data makes it possible for a measurement of $\delta_{\rm CP}$ with
an error of $\pm 10^\circ$ at $\delta_{\rm CP}=90^\circ$.

Actually, it is pointed out that all three types of neutrino beams have the discovery potential for the CP violation given the relatively large $\theta_{13}$~\cite{Coloma:2012wq}. Though the $3\sigma$ discovery region are limited to $25\%$ of all possible values for $\delta_{\rm CP}$ in the upgraded T2K and NOvA experiments, the overall $3\sigma$ discovery reach of the Long-Baseline Neutrino Experiment (LBNE) can be around $70\%$ of all possible values for $\delta_{\rm CP}$~\cite{Akiri:2011dv}. Simulations~\cite{Bishai:2012ss} have shown that our prediction of $\delta_{\rm CP}$ lies in the range that can be directly examined in the Project-X of the LBNE.

It is interesting to notice that our procedure leads to a
quasi-maximal CP violation in the standard parametrization, and such
prediction differs from some theoretical
expectations~\cite{Xing:2012ej,Wu:2012ri,Meloni:2012sy}. However,
there have been some theoretical investigations indicating that a
large CP violating phase $\delta_{CK}$ can be understood from some
basic asymmetries. The near maximal CP violation with a large
$\theta_{13}$ from our analysis is in accordance with a general
approach based on residual $Z_2$ symmetries~\cite{Ge:2011qn}. A
maximal CP violation is also predicted from the octahedral symmetry
for the family symmetry of the neutrino-lepton
sector~\cite{He:2012yt}. In Ref.~\cite{Hernandez:2012ra}, a
prediction of $\delta_{\rm CK}=(60-90)^\circ$ is made within the
framework of discrete groups, i.e. $A_4$, $S_4$ and $A_5$.
Actually, the range for $\delta_{\rm CK}=(60-90)^\circ$ can be
translated into $\delta_{\rm KM}=(60.31-90)^\circ$ by noticing
\begin{eqnarray}
\sin\delta_{\rm
KM}=\frac{\cos\theta_{13}\sin2\theta_{12}\sin2\theta_{23}\sin2\theta_{13}}{\sin\theta_1\sin2\theta_1\sin2\theta_2\sin2\theta_3}\sin\delta_{\rm
CK}=1.00311 \sin\delta_{\rm CK}.
\end{eqnarray}
Thus our prediction of a quasi-maximal $\delta_{\rm CK}$ or a
maximal $\delta_{\rm KM}$ can acquire the theoretical support from
basic considerations.

In summary, we can predict the neutrino mixing matrix with all
elements determined together with a prediction of the CP violating
phase. We also predict the three angles of the unitarity triangle
corresponding to the quark sector for confronting with the
CP-violation related measurements. A similar exercise can be
performed to the quark case, and we can get proved that the same
procedure can also lead to a successful reproduction of the CP
violating phase $\delta^{\mathrm{quark}}_{\mathrm{CK}}$ in the
standard parametrization of the CKM mixing matrix. Our prediction is
model independent without any ambiguity, except that the parameters
can be also gotten from global fitting procedure instead of the
analytic expressions adopted in this paper. We expect a test of our
prediction of the full neutrino mixing matrix and the corresponding
CP violating phase, or the three angles of the unitarity triangle in
a convention independent manner as in the quark case, through future
experiments.

We acknowledge discussions with Nan Qin and Ya-juan Zheng.
This work is partially supported by National Natural
Science Foundation of China (Grants No.~11021092, No.~10975003,
No.~11035003, and No.~11120101004) and by the Research Fund for the
Doctoral Program of Higher Education (China).

\end{document}